\documentclass[12pt,preprint]{aastex}

\shorttitle{Exoplanet Host Star Diameters}
\shortauthors{Baines et al.}

\begin{document}

\title{Eleven Exoplanet Host Star Angular Diameters \\ from the CHARA Array}

\author{Ellyn K. Baines$^\dagger$}
\affil{Remote Sensing Division, Naval Research Laboratory, 4555 Overlook Avenue SW, Washington, DC 20375}
\email{ellyn.baines.ctr@nrl.navy.mil}
\altaffiltext{$^\dagger$}{The observations described here were completed while with the Center for High Angular Resolution Astronomy, Georgia State University, P.O. Box 3969, Atlanta, GA 30302-3969.}

\author{Harold A. McAlister, Theo A. ten Brummelaar, Judit Sturmann, Laszlo Sturmann, \& Nils~H.~Turner}
\affil{Center for High Angular Resolution Astronomy, Georgia State University, P.O. Box 3969, Atlanta, GA 30302-3969}
\email{hal@chara.gsu.edu; theo, judit, sturmann, nils@chara-array.org}

\author{Stephen T. Ridgway}
\affil{Kitt Peak National Observatory, National Optical Astronomy Observatory, \\ P.O. Box 26732, Tucson, AZ 85726-6732} 
\email{ridgway@noao.edu}

\altaffiltext{}{For preprints, please email ellyn.baines.ctr@nrl.navy.mil.}

\begin{abstract}
We directly measured the angular diameters for eleven exoplanet host stars using Georgia State University's CHARA Array interferometer and calculated their linear radii and effective temperatures. The sample tends towards evolving or evolved stars and includes one dwarf, four subgiants, and six giants. We then estimated masses and ages for the stars using our effective temperatures combined with metallicity measurements from the literature.
\end{abstract}


\section{Introduction}
Exoplanets are discovered on a regular basis, most via radial velocity surveys and transiting events. Many host star angular diameters have been estimated using photometric and spectroscopic methods \citep[e.g.,][respectively]{2003A&A...411L.501R,2005ApJ...622.1102F}, and while these are excellent for approximating angular diameters, they are by nature indirect methods. The advantage interferometry brings is the ability to directly measure the angular sizes of the stars, which in turn leads to physical radii and effective temperatures. These are important parameters that describe the parent star as well as the environment in which the exoplanet resides.

This paper represents an extension and continuation of the work described in \citet{2008ApJ...680..728B}, where the angular diameters for 24 exoplanet host stars were published. While the previous sample featured a few giants and some subgiants, well over half were dwarfs or stars showing signs of just beginning to evolve off the main-sequence. This paper focuses on giants and subgiants, and only one dwarf is represented.

%%%%%%%%%%%%%%%%%%%%%%%%%%%%%%%%%%%% Observations %%%%%%%%%%%%%%%%%%%%%%%%%%%%%%%%%%%%%%%%%%%%%%
\section{Interferometric Observations}
All observations were obtained using the Center for High Angular Resolution Astronomy (CHARA) Array, a six-element optical/infrared interferometric array located on Mount Wilson, California \citep{2005ApJ...628..453T}. We used the pupil-plane ``CHARA Classic'' beam combiner in the $K'$-band (2.15~$\mu$m), paired with the longest baseline the Array offers at 331~m. The observing procedure and data reduction process employed here are described in \citet{2005ApJ...628..439M}. Table \ref{observations} lists the exoplanet host stars observed, their calibrators, the dates of the observations, and the number of observations obtained.

Our target list was culled from the complete exoplanet list by using declination limits and magnitude constraints: north of -10$^\circ$ declination, brighter than $V=+10$ in order for the tip/tilt system to lock onto the star, and brighter than $K=+6.5$ so fringes were easily visible. We obtained data on the 11 exoplanet host stars over two observing runs in July and September 2008.

\begin{figure}[!h]
  \centering \includegraphics[angle=90,width=1.0\textwidth]
  {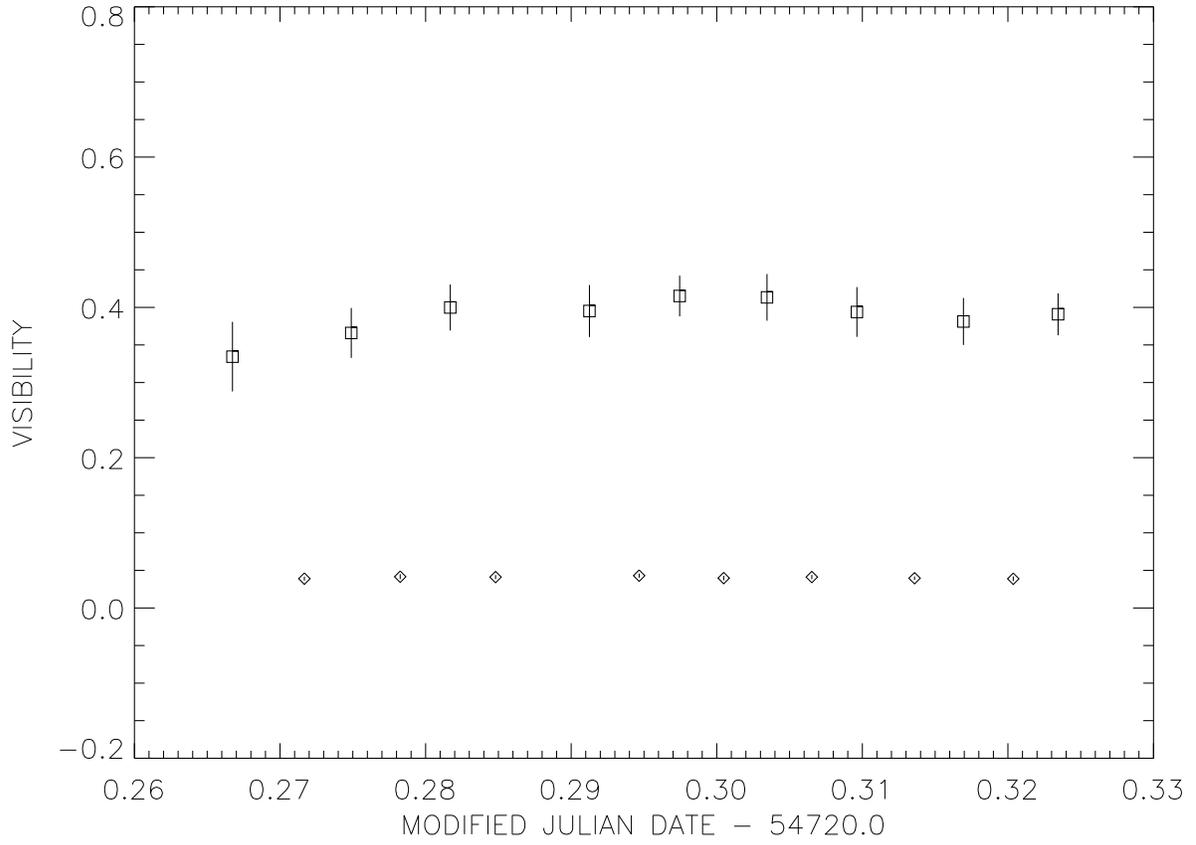}\\
  \caption{Uncalibrated visibilities for HD~222404 from 2008/09/11. The squares and diamonds are the calibrator's and target's measured visibilities, respectively, and the vertical lines are the errors in those visibilities.}
  \label{uncalib_data}
\end{figure}

Reliable calibrators stars are critical in interferometric observations, acting as the standard against which the science target is measured, so every effort was made to find spherical, non-variable, single-star calibrators. Our observing pattern was calibrator-target-calibrator so that every target was flanked by calibrator observations made as close in time as possible; therefore ``10 bracketed observations'' denotes 10 object and 11 calibrator data sets, each of which is comprised of approximately 200 scans across the fringe. This allowed us to calculate the target's calibrated visibilities from the instrumental visibilities of the target and calibrator. Figure~\ref{uncalib_data} shows an example of uncalibrated visibilities. Acceptable calibrators were chosen to be smaller than $\sim$0.4 milliarcseconds (mas), so they were nearly unresolved and uncertainties in their diameters did not affect the target's diameter calculation as much as if the calibrator had a significant angular size on the sky.

In order to estimate the calibrator stars' angular diameters as well as check for excess emission that could indicate a low-mass stellar companion or circumstellar disk, we fitted spectral energy distributions (SEDs) based on published $UBVRIJHK$ photometric values for each star. Limb-darkened diameters were calculated using Kurucz model atmospheres\footnote{See http://kurucz.cfa.harvard.edu.} based on effective temperature ($T_{\rm eff}$) and gravity (log~$g$) values obtained from the literature. The models were then fit to observed photometric values also from the literature after converting magnitudes to fluxes using \citet{1996AJ....112..307C} for $UBVRI$ values and \citet{2003AJ....126.1090C} for $JHK$ values. 

Table \ref{observations} lists the $T_{\rm eff}$ and log~$g$ used for each calibrator, the resulting limb-darkened angular diameters, and the distance between the target and calibrator stars. We used calibrators as close to the target star as possible. The target-calibrator (T-C) distances ranged from 1 to 9$^\circ$ and all but two calibrators were within 5$^\circ$ of their target stars. This allowed us to observe the stars as close together in time as possible, usually on the order of 3 to 5 minutes between the two. For the T-C pairs of 8 and 9$^\circ$, the slightly greater distance added little to the error in the diameter measurement. Table~\ref{calib_info} provides more details on each calibrator star used, and Table~\ref{calib_visy} lists the Modified Julian Date (MJD), projected baseline ($B$), projected baseline position angle ($\Theta$), calibrated visibility ($V_{\rm c}$), and error in $V_c$ ($\sigma V_{\rm c}$) for each exoplanet host star observed.

%%%%%%%%%%%%%%%%%%%%%%%%%%%%%%% Diameter Determination %%%%%%%%%%%%%%%%%%%%%%%%%%%%%%%%%%
\section{Angular Diameter Determinations}
\begin{figure}[!h]
  \centering \includegraphics[width=1.0\textwidth]
  {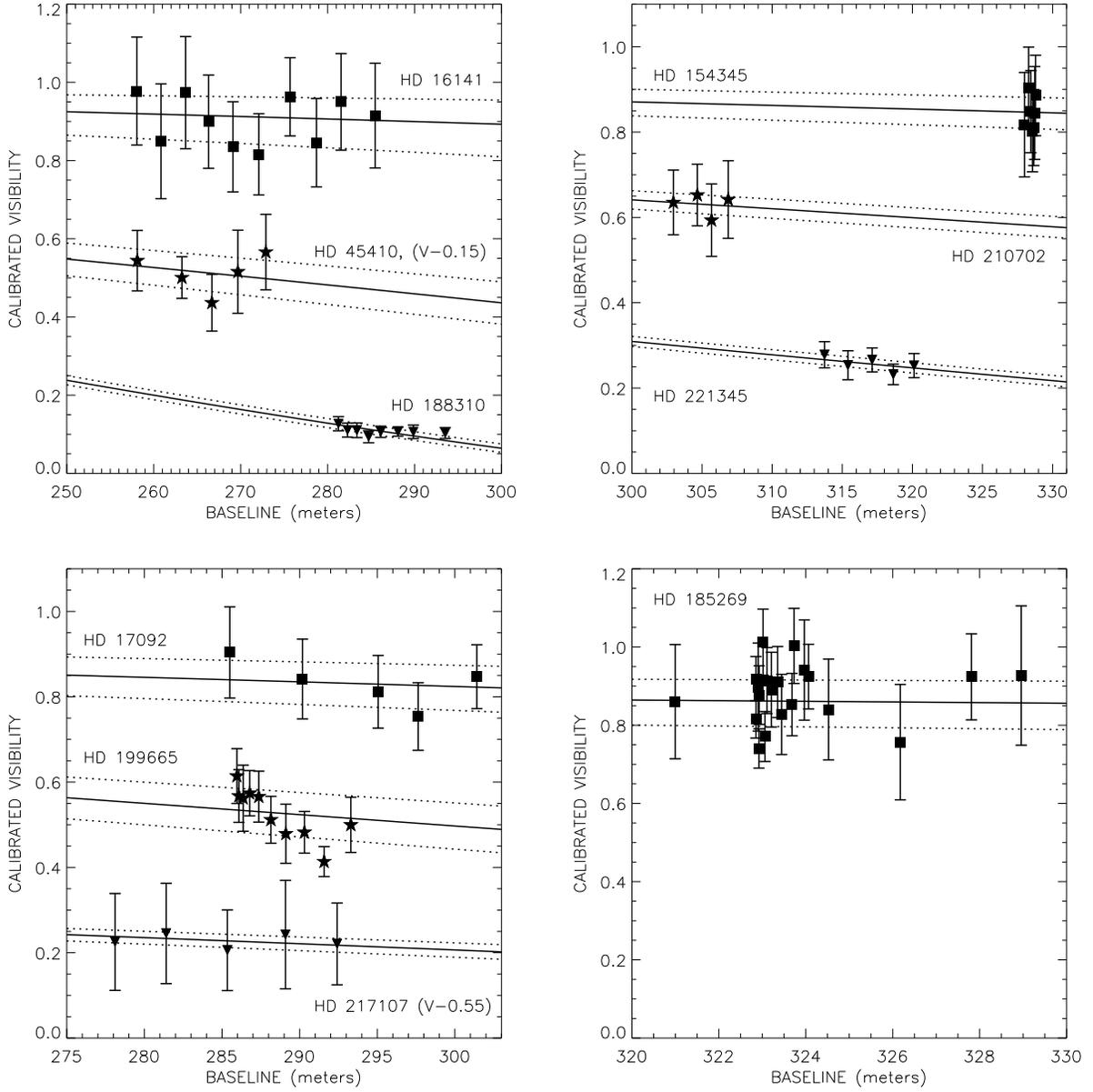}\\
  \caption{LD disk diameter fits for all the stars except HD 222404. The solid lines represent the theoretical visibility curve with the best fit $\theta_{\rm LD}$ for each star, the dashed lines are the 1$\sigma$ error limits of the diameter fit, the solid symbols are the calibrated visibilities, and the vertical lines are the measured errors. HD 45410's and HD 217107's visibilities were subtracted by the offset indicated by ``(V - $\#$)'' so they would not overlap other data points.}
  \label{lddiam_all}
\end{figure}

\begin{figure}[!h]
  \centering \includegraphics[width=0.55\textwidth]
  {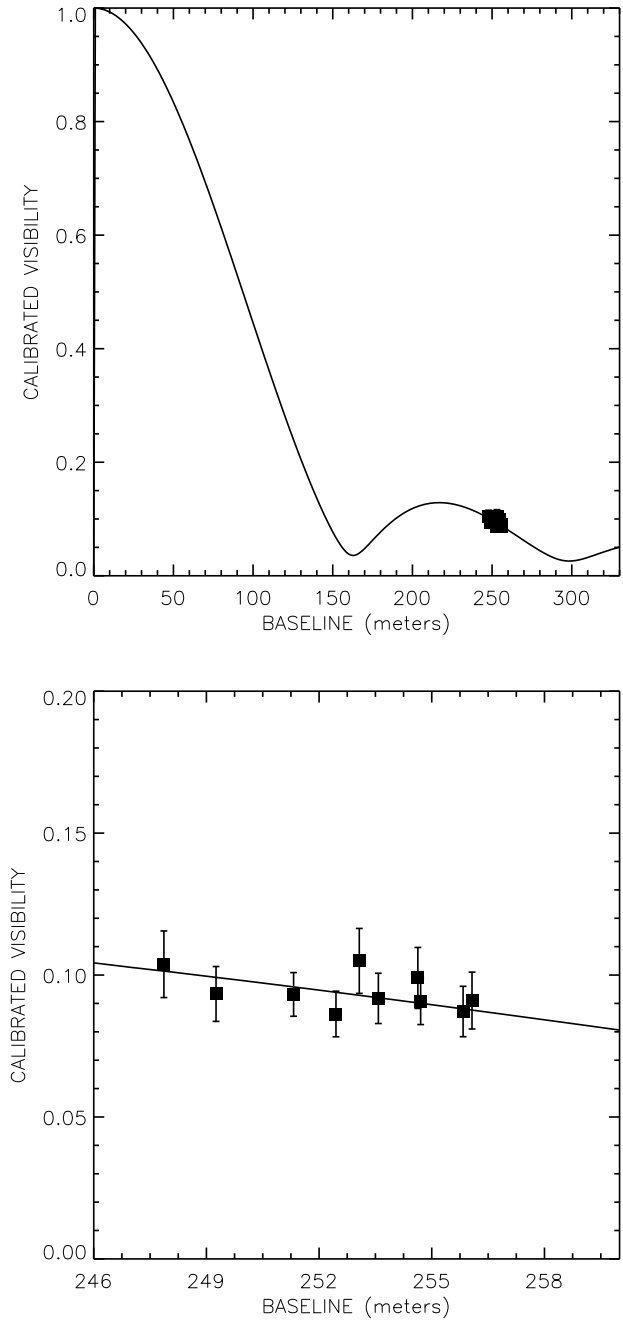}\\
  \caption{LD disk diameter fit for HD 222404. The solid line represents the theoretical visibility curve for the best fit $\theta_{\rm LD}$, the squares are the calibrated visibilities, and the vertical lines are the measured errors. The top panel shows the full visibility curve with the 10 data points clustered on the second lobe, and the bottom panel zooms in on those data points.}
  \label{lddiam_HD222404}
\end{figure}

Diameter fits to visibilities ($V$) were based upon the uniform disk (UD) approximation given by $V = [2 J_1(x)] / x$, where $J_1$ is the first-order Bessel function and $x = \pi B \theta_{\rm UD} \lambda^{-1}$, where $B$ is the projected baseline at the star's position, $\theta_{\rm UD}$ is the apparent UD angular diameter of the star, and $\lambda$ is the effective wavelength of the observation \citep{1992ARAandA..30..457S}. The limb-darkened (LD) relationship incorporating the linear limb darkening coefficient $\mu_{\lambda}$ \citep{1974MNRAS.167..475H} is:
\begin{equation}
V = \left( {1-\mu_\lambda \over 2} + {\mu_\lambda \over 3} \right)^{-1}
\times
\left[(1-\mu_\lambda) {J_1(\rm x) \over \rm x} + \mu_\lambda {\left( \frac{\pi}{2} \right)^{1/2} \frac{J_{3/2}(\rm x)}{\rm x^{3/2}}} \right] .
\end{equation}
Figures~\ref{lddiam_all} and \ref{lddiam_HD222404} show the LD diameter fits for all the stars. Though the difference between LD and UD diameters is a minor effect in the wavelength used here, the former have the advantage over the latter in that they are better suited to calculating effective temperatures and more closely represent the physical properties of the star \citep{2009ApJ...694.1085V}. 

For each $\theta_{\rm LD}$ fit, the errors were derived via the reduced $\chi^2$ minimization method: the diameter fit with the lowest $\chi^2$ was found and the corresponding diameter provided the final $\theta_{\rm LD}$ for the star. The errors were calculated by finding the diameter at $\chi^2 + 1$ on either side of the $\chi^2_{\rm min}$ and determining the difference between the $\chi^2$ diameter and $\chi^2 +1$ diameter.

Our experience has shown that the rms of the residuals to diameter fits of visibilities is typically smaller than the mean of the standard errors attributed to each contributing visibility measurement. As described by \citet{2005ApJ...628..439M}, the error estimates assigned to calibrated visibilities were determined by the rms of the means of subsets of the entire sample of visibility measurements made at a particular epoch. We now find that this approach tends to overestimate the error of individual visibilities, producing reduced $\chi^2$ values well under 1.0. This, in turn, leads to overestimates of the errors in angular diameter. In calculating the diameter errors in Table~\ref{calculating_diam}, we have adjusted the estimated visibility errors by a factor that forces the reduced $\chi^2$ to unity, and we believe the resulting diameter errors are more representative of the influence of the true intrinsic errors in our visibilities.

Table \ref{calculating_diam} lists the following parameters for each star: spectral type, $\mu_{\lambda}$, the \emph{Hipparcos} parallax \citep[$\pi$,][]{2007hnrr.book.....V}, the LD diameter estimated from SED fits ($\theta_{\rm SED}$), the UD and LD angular diameters $\theta_{\rm UD}$ and $\theta_{\rm LD}$, and the linear radius ($R_{\rm L}$) derived from the combination of $\theta_{\rm LD}$ and $\pi$. Six of the stars had $\theta_{\rm SED}$ calculated by \citet{2009ApJ...694.1085V}, and Table~\ref{calculating_diam} lists the photometric sources for the remaining stars, whose SED fits were completed by us as described in $\S$2. The $T_{\rm eff}$ and log~$g$ values used in our SED fits were from \citet{1999A&A...352..555A} for all the stars except HD~17092 and HD~154345, which were from \citet{2000asqu.book.....C} and \citet{2005ApJS..159..141V}, respectively. The star HD~17092 does not have any available parallax measurements, so we used the photometric distance estimate from \citet{2008AstL...34..785G} with an assigned error of 10$\%$.

\begin{figure}[!h]
  \centering \includegraphics[width=1.0\textwidth]
  {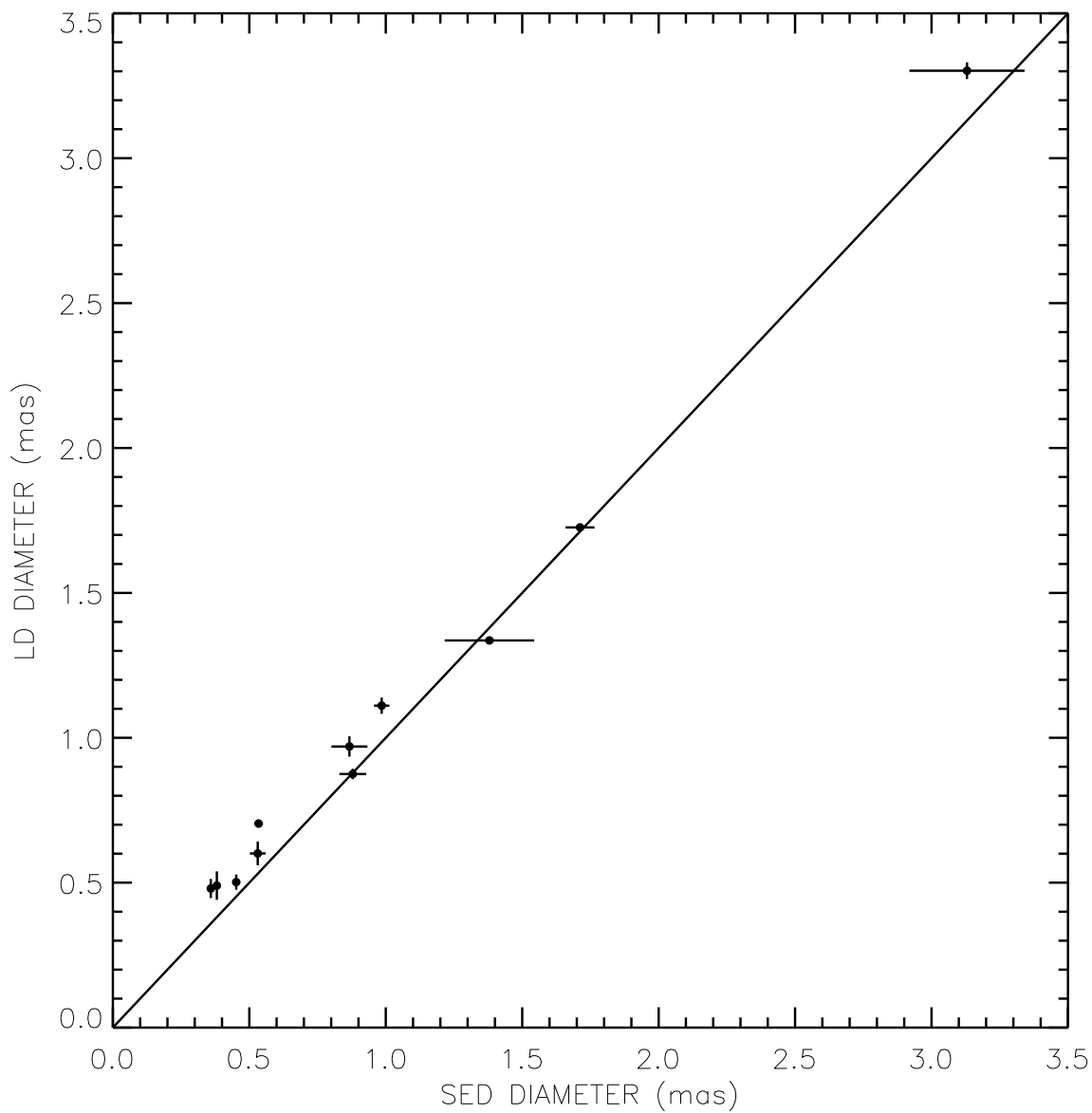}\\
  \caption{A comparison of estimated SED diameters and measured LD diameters with their corresponding errors. The solid line indicates a 1:1 ratio for the diameters. The LD diameter errors are consistently low, ranging between 0.01 to 0.05 mas, while the SED diameter errors show a wider spread, from 0.01 to 0.16 mas, and are dependent on how well the stellar model's fluxes match the measured values. In the case of HD~221345 and HD~222404, which are the two points showing the largest SED errors, the model fluxes do not correspond as well to the measured fluxes.}
  \label{sedld}
\end{figure}

To check how well the estimated angular diameters match the measured values, Figure~\ref{sedld} plots $\theta_{\rm SED}$ versus $\theta_{\rm LD}$ and shows how the SED diameters slightly underestimate the true sizes of these evolved stars. This may be due to model assumptions about opacity that are not exactly true to life.

%Measuring angular diameters using this method involve two main assumptions. First, that the exoplanet host star does not have any unseen stellar companions that would alter the measured visibilities. A companion check was performed for each star by studying any possible systematics in the single-star uniform-disk fit errors. Second, that the calibrator star's angular diameter was known and could be used to calibrate the target star's visibilities. If the calibrated visibilities exceeded 1 for a given dataset, that calibrator was discarded and the target star was observed again with a new calibrator.

Two stars have been previously measured interferometrically: HD 221345 and HD 222404. \citet{1999AJ....117..521V} observed HD 221345 using the Palomar Testbed Interferometer \citep{1999ApJ...510..505C} and their value of $\theta_{\rm UD}$ was 1.75$\pm$0.07 mas. \citet{1999AJ....118.3032N} used the Navy Prototype Optical Interferometer \citep[NPOI,][]{1998ApJ...496..550A} to measure HD 222404 and their LD angular diameter of 3.24$\pm$0.03 mas is close to our measurement of 3.30$\pm$0.01 mas. The NPOI observes in visible wavelengths and therefore the limb-darkening effects will be larger and more model dependent than is the case for data from the CHARA Array.

Our data for HD~222404 are on the second lobe of the visibility curve (Figure \ref{lddiam_HD222404}), and the second lobe is where second-order effects such as limb darkening start to have more of an influence than on the first lobe. In order to check that we are fitting only the angular diameter to these data and are not making unfair assumptions about the limb darkening coefficient, we determined the diameter after changing $\mu_{\lambda}$ by 50$\%$, which is well past the regime for stars of HD~222404's general $T_{\rm eff}$ and log~$g$. The resulting change in diameters was $\sim 0.6\%$, indicating a low dependence on the $\mu_{\lambda}$ used.

%The difference between their diameter value and ours of 1.38$\pm$0.01 mas is likely due to the nature of the baseline available. van Belle et al. used a 109-m baseline while we used a 331-m baseline, and our visibilities are significantly closer to the first null than those obtained using PTI ($V_{\rm CHARA}$ = 0.23-0.28 versus $V_{\rm PTI}$ = 0.78-0.81), indicating a higher sensitivity to stellar diameter measurements.
%Nordgren et al. used baselines up to 37.5 m, and, again, our observations at 331 m allow us to be more accurate. Though the NPOI's baseline was significantly shorter, they observed in visible wavelengths so the difference in the diameter accuracy is not as acute as it would be if they observed using a 37.5-meter baseline in the $K$-band.

Many of the stars in the sample are published in the literature as variable stars or as components in a binary star system. Table \ref{binvar} lists the stars, the pertinent references, and why their variability or binarity do not affect our measurements here. For the variable stars, no reliable periods or types are listed in the literature, and if those stars are variable, it is on a level not likely to have a significant impact on our measurements. As for the binary star systems, the companions are too far away from the primary star and well out of the field of view (FOV) of the CHARA Array and/or the magnitude difference is too great for the Array to detect the secondary star. 

The range of binary separations available to the CHARA Array, taking all the baselines into account, is approximately 10 mas to 1.0 arcsecond, while the maximum FOV of the baseline used for our observations is $\sim$230 mas. The lower limit of binary detection using the CHARA Array is 2.5 magnitudes in the $K$-band, and this value depends on the absolute brightness of the two stars and could therefore be higher for some systems. It is possible that the exoplanet parent stars may also host low-mass stellar companions not detected by the Array, though it is more likely they would have been detected by the radial velocity studies. We cannot detect the exoplanets themselves using the Array, due to the large magnitude difference between star and planet.

%%%%%%%%%%%%%%%%%%%%%%%%%%%%%%%%%%%% Calculating Teff %%%%%%%%%%%%%%%%%%%%%%%%%%%%%%%%%%%%%%%%%%%%

\section{Effective Temperatures}
Once $\theta_{\rm LD}$ is measured, the effective temperature can be calculated using the relation 
\begin{equation}
F_{\rm BOL} = {1 \over 4} \theta_{\rm LD}^2 \sigma T_{\rm eff}^4,
\end{equation}
where $F_{\rm BOL}$ is the bolometric flux and $\sigma$ is the Stefan-Bolzmann constant. $F_{\rm BOL}$ was determined by applying the bolometric corrections (BC) for each star after taking interstellar absorption ($A_{\rm V}$) into account. Table \ref{temps} lists the $A_{\rm V}$ and BC used, and the resulting $F_{\rm BOL}$ and $T_{\rm eff}$. As a comparison, a range of $T_{\rm eff}$ from other sources is also listed in Table \ref{temps}. Five stars have $T_{\rm eff}$ within their ranges of temperatures obtained using other means, five stars are slightly out of their ranges but are within measured errors, and only one star is significantly outside its range (HD~185269, by $\sim$570 K). This could be due to incorrect spectral typing or assumptions about factors such as opacity and metallicity that are buried in the model used for each of the three references that list temperatures for this star.

Because the $\theta_{\rm LD}$ is dependent on the $\mu_\lambda$ value selected, which in turn is dependent on log~$g$ and $T_{\rm eff}$, we wanted to check the effect of the new temperature values on measured LD diameters. Using the newly-calculated $T_{\rm eff}$ to find $\mu_{\lambda}$, we found the average difference in $\mu_{\lambda}$ was $<$6$\%$ and the resulting $\theta_{\rm LD}$ values differed on average of 0.3$\%$, indicating this is a negligible effect.

%%%%%%%%%%%%%%%%%%%%%%%%%%%%%%%%%%%% Stellar Models %%%%%%%%%%%%%%%%%%%%%%%%%%%%%%%%%%%%%%%%%%%%

\section{Stellar Model Results}
%In order to determine stellar ages, we used the CMD 2.1 model\footnote{http://stev.oapd.inaf.it/cgi-bin/cmd} in the Padova database of stellar evolutionary tracks and isochrones, which provides interpolated isochrones for a wide range of metallicities \citep{2008A&A...482..883M}. The model outputs of interest to us were stellar luminosity and temperature, as the luminosity could be calculated from the star's bolometric magnitude and $T_{\rm eff}$ is now known from Equation 3. 

%For each star, the metallicity from a range of sources from \citet{2000A&AS..143...23O} were averaged and then used as input for the model. We plotted log($L_{\odot}$) versus log($T_{\rm eff}$) for each star with a variety of isochrones of different ages in Figures 3 through 12. Table \ref{temps} lists the luminosities used and Table \ref{models} lists the resulting age, if available. For some stars the errors in $T_{\rm eff}$ are too large to constrain ages, in others the placement of the data point is inconclusive. In the case of HD~217107, the star's metallicity was higher than the model allowed, which accounts for the isochrones' separation from the data point.

In order to estimate stellar ages, masses, and linear radii, we used the PARAM 1.0 model\footnote{http://stev.oapd.inaf.it/cgi-bin/param$\_$1.0} \citep{2006A&A...458..609D}, which is based on a set of theoretical isochrones from \citet{2000A&AS..141..371G}. The model uses each star's metallicity, effective temperature, and $V$ magnitude to estimate its age, mass, radius, $(B-V)_0$, and log~$g$ using the isochrones and a Bayesian estimating method, calculating the probability density function separately for each property in question. da Silva et al. are most confident in resulting $(B-V)_0$, log~$g$, radii, and angular diameter predictions while describing the age and mass estimates as ``more uncertain''. We left the Bayesian priors (initial mass function and star formation rate in a given interval) on the default settings when running the model.

The model's inputs were the star's $T_{\rm eff}$, [Fe/H], $V$ magnitude, and parallax along with the corresponding error for each value. $T_{\rm eff}$ was calculated using Equation 3, the $V$ magnitude was from Mermilliod (1991), the parallax was from \citet{2007hnrr.book.....V}, and the [Fe/H] value was averaged from all the sources available from \citet{2000A&AS..143...23O} with its error represented by the standard deviation of all the measurements. When only one source of [Fe/H] was in the literature (the case for HD 45410 and HD 185269), an error of 0.05 was assigned. The same error was used when the star had no [Fe/H] listed and solar metallicity was assumed (the case for HD 17092, HD 154345, and HD 210702).

\begin{figure}[!h]
  \centering \includegraphics[width=1.0\textwidth]
  {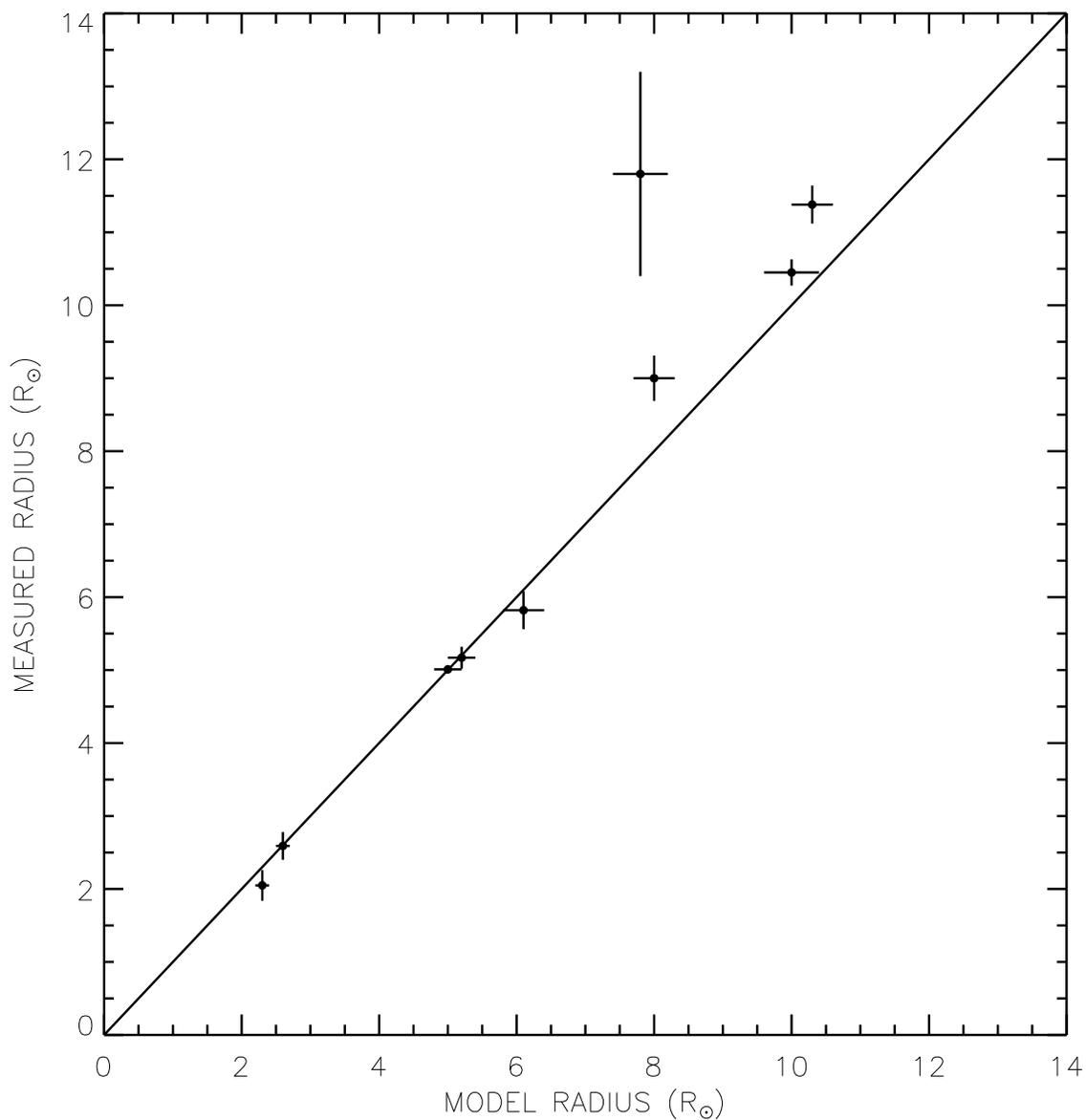}\\
  \caption{A comparison of model and measured radii with their corresponding errors. The solid line indicates a 1:1 ratio for the radii. The measured radii errors depend on uncertainties in the LD diameter and parallax measurements while the model radii errors depend on the model's inputs, including effective temperature, metallicity, and parallax measurements. The errors in each input value contribute to the error budget of the model radius. The largest outlier is HD~17092, which had the least reliable distance measurement of the sample.}
  \label{radii}
\end{figure}

\begin{figure}[!h]
  \centering \includegraphics[angle=90,width=1.0\textwidth]
  {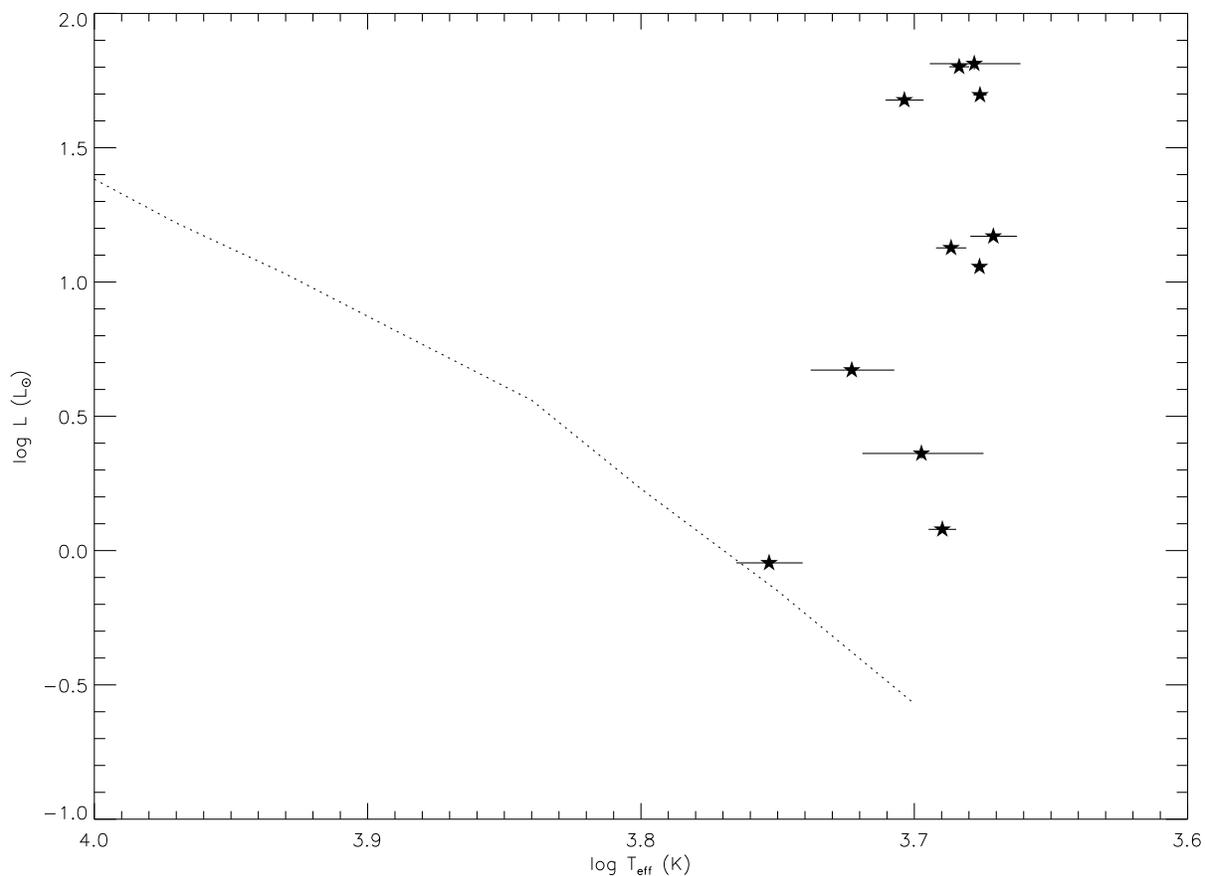}\\
  \caption{H-R diagram for the exoplanet host stars. The dotted line indicates the ZAMS derived from \citet{2000asqu.book.....C}. The star closest to this line is HD~154345 and is the only dwarf in the sample. The remaining points represent the giant branch of the H-R diagram. The main sources of error in the luminosity values arise from uncertainties in bolometric corrections (the error bars are within the data points), while the effective temperature errors depend on uncertainties in the star's parallax and LD diameter measurements.}
  \label{hr}
\end{figure}

The resulting age, mass, and $R_{\rm model}$ are listed in Table \ref{models} for all the stars except HD 154345 because it is a dwarf and the model is for evolving stars, and for HD 217107, whose metallicity is out of range of the model. Figure \ref{radii} plots the model's radii versus those measured interferometrically. The agreement between the two is excellent for the small to intermediate-sized stars, but the model appears to systematically underestimate the radii for the four largest stars. Figure \ref{hr} plots luminosity versus $T_{\rm eff}$ and  represents the Hertzsprung-Russell (H-R) diagram. The zero-age main-sequence (ZAMS) line is shown as derived from \citet{2000asqu.book.....C} and the one dwarf in the sample (HD~154345) is the point nearly on the ZAMS while the other stars form the giant branch.

%%%%%%%%%%%%%%%%%%%%%%%%%%%%%%%%%%%%% Conclusion %%%%%%%%%%%%%%%%%%%%%%%%%%%%%%%%%%%%%%%%%%%%%%%%%
\section{Conclusion}
We measured the angular diameters of 11 exoplanet host stars for a sample almost entirely comprised of evolving and evolved stars. All LD diameters boasted errors of $\leq$10$\%$, and 8 of the 11 had errors $\leq$5$\%$. Linear radii were derived from $\theta_{\rm LD}$ and the stars' \emph{Hipparcos} measurements, and we calculated effective temperatures using our $\theta_{\rm LD}$ values. The subsequent errors on the $T_{\rm eff}$ were all $\leq$5$\%$. 

Using our new effective temperatures, [Fe/H] values from the literature, and the PARAM stellar model, we were able to estimate the radii, masses, and ages for the stars, and the model radii match the measured radii well for all the giants except the four largest stars in the sample. Previous interferometric measurements of other giant stars showed a similar effect, where high-luminosity stars have larger radii at a given effective temperature \citep{1998AJ....116..981D}. The four stars in question - HD~17092, HD~188310, HD~199665, and HD~221345 - are by far the most luminous stars in the sample so it is not entirely unexpected that the models underestimate their radii. It would be to the model's advantage if it could be modified to incorporate this effect.

By directly measuring exoplanet host stars' angular diameters and calculating the physical radii and temperatures, we are able to better characterize the exoplanets' environments. We now know that solar systems come in many different configurations \citep{2006ApJ...646..505B}, and interferometric measurements help to describe the all-important central stars. This in turn will help to constrain parameters such as the location and size of the habitable zone as well as putting limitations on the temperature profiles of the planets themselves.

%%%%%%%%%%%%%%%%%%%%%%%%%%%%%%%%%% Acknowledgements %%%%%%%%%%%%%%%%%%%%%%%%%%%%%%%%%%%%%%%%%%%%%%

\acknowledgements

Many thanks to P.J. Goldfinger and Chris Farrington for their invaluable assistance in obtaining the data used here, and to Tabetha Boyajian for her very helpful suggestions. The CHARA Array is funded by the National Science Foundation through the NSF grant AST-0606958 and by Georgia State University through the College of Arts and Sciences. This research has made use of the SIMBAD literature database, operated at CDS, Strasbourg, France, and of NASA's Astrophysics Data System. This publication also makes use of data products from the Two Micron All Sky Survey, which is a joint project of the University of Massachusetts and the Infrared Processing and Analysis Center/California Institute of Technology, funded by the National Aeronautics and Space Administration and the National Science Foundation.

%%%%%%%%%%%%%%%%%%%%%%%%%%%%%%%%%%%%%%%%%%%%%%%%%%%%%%%%%%%%%%%%%%%%%%%%%%%%%%%

\clearpage

%%%%%%%%%%%%%%%%%%%%%%%%%%%%%%%%%%% Observing Log %%%%%%%%%%%%%%%%%%%%%%%%%%%%%%%%%%%%%%%%%%%%

\begin{deluxetable}{cccccccccc}
\tabletypesize{\scriptsize}
\tablewidth{0pc}
\tablecaption{Observing Log and Calibrator Stars' Basic Parameters.\label{observations}}

\tablehead{\multicolumn{5}{c}{Observing Log} & \multicolumn{1}{c}{ } & \multicolumn{3}{c}{Calibrator Information} \\
\cline{1-5} \cline{7-9} \\
 \colhead{Target} & \colhead{Other} & \colhead{Calibrator} & \colhead{Date} & \colhead{\# Bracketed}    & \colhead{ } & \colhead{$T_{\rm eff}$} & \colhead{log~$g$}      & \colhead{$\theta_{\rm LD,SED}$} & \colhead{T-C Sep} \\
  \colhead{HD}    & \colhead{Name}  & \colhead{HD}         & \colhead{(UT)} & \colhead{Observations} & \colhead{ } & \colhead{(K)}           & \colhead{(cm s$^{-2}$)} & \colhead{(mas)} & \colhead{(deg)} \\ }
\startdata
16141  & 79 Cet   & 18331  & 2008/09/09 & 10 & & 8710 & 4.14 & 0.354$\pm$0.019 & 5 \\
17092  & $\ldots$ & 14212  & 2008/09/11 & 5  & & 9333 & 4.08 & 0.291$\pm$0.006 & 5 \\
45410  & 6 Lyn    & 46590  & 2008/09/11 & 5  & & 9550 & 4.14 & 0.221$\pm$0.007 & 2 \\
154345 & $\ldots$ & 151044 & 2008/09/10 & 7  & & 6166 & 4.38 & 0.380$\pm$0.008 & 4 \\
185269 & $\ldots$ & 184381 & 2008/07/18 & 15 & & 6650 & 4.34 & 0.285$\pm$0.010 & 3 \\
       &          &        & 2008/07/20 & 5  & & $\ldots$ & $\ldots$ & $\ldots$ & $\ldots$ \\
188310 & $\xi$ Aql & 182101 & 2008/09/08 & 8 & & 6607 & 4.33 & 0.344$\pm$0.014 & 8 \\
199665 & 18 Del   & 194012 & 2008/09/08 & 10 & & 6310 & 4.36 & 0.441$\pm$0.016 & 9 \\
210702 & $\ldots$ & 210074 & 2008/09/08 & 4  & & 7079 & 3.82 & 0.384$\pm$0.013 & 4 \\
217107 & $\ldots$ & 217131 & 2008/09/08 & 5  & & 6918 & 3.71 & 0.305$\pm$0.014 & 1 \\
221345 & 14 And   & 222451 & 2008/09/11 & 5  & & 6761 & 4.22 & 0.346$\pm$0.011 & 3 \\
222404 & $\gamma$ Cep & 219485 & 2008/07/17 & 3  & & 9790 & 4.14 & 0.214$\pm$0.006 & 4 \\
       &              &        & 2008/09/11 & 7  & & $\ldots$ & $\ldots$ & $\ldots$ & $\ldots$ \\
\enddata
\tablecomments{$T_{\rm eff}$ and log~$g$ values come from Allende Prieto \& Lambert (1999), except for HD~184381 and HD~219485, whose $T_{\rm eff}$ and log~$g$ values are based on spectral type as listed in the \emph{SIMBAD Astronomical Database} and \citet{2000asqu.book.....C}. }
\end{deluxetable}

\clearpage

%%%%%%%%%%%%%%%%%%%%%%%%%%%%%%%%%%% Calib Details %%%%%%%%%%%%%%%%%%%%%%%%%%%%%%%%%%%%%%%%%%%%

\begin{deluxetable}{cl}
\tablewidth{0pc}
\tablecaption{Previous Calibrator Uses.\label{calib_info}}
\tablehead{\colhead{Calib HD} & \colhead{ } }
\startdata
14212  & Used as calibrator in \citet{2009ApJ...694.1085V} \\
18331  & Used as calibrator in \citet{2009ApJ...694.1085V} \\
46590  & Considered a single star in Royer et al. (2007) \\
151044 & Used as calibrator in \citet{2008ApJ...680..728B} \\
182101 & Used as calibrator in \citet{2006ApJ...644..475B} \\
184381 & Used as calibrator in \citet{2006ApJ...652.1724J} \\
194012 & Used as calibrator in \citet{2008ApJ...680..728B} \& \citet{Montes}; \\
       & no binary companion found in \citet{1987AJ.....93..183M} \\
210074 & Used as comparison star in \citet{2005ApJ...632.1157W} \& \citet{2000ApJ...529L..41H} \\
217131 & Used as comparison star in \citet{2005ApJ...632..638V}; \\
       & no binary companion found in \citet{1987AJ.....93..183M} \\
219485 & Considered a single star in Royer et al. (2007) \\
222451 & Considered a single star in Nordstr{\"o}m et al. (2004) \\

\enddata
\end{deluxetable}

\clearpage

%%%%%%%%%%%%%%%%%%%%%%%%%%%%% All Calibrated Visibilities %%%%%%%%%%%%%%%%%%%%%%%%%%%%%%%%%%%

\begin{deluxetable}{cccccc}
\tablewidth{0pc}
\tablecaption{Calibrated Visibilities.\label{calib_visy}}

\tablehead{
 \colhead{Target} & \colhead{ }   & \colhead{$B$} & \colhead{$\Theta$} & \colhead{ }     & \colhead{ } \\
 \colhead{HD}   & \colhead{MJD} & \colhead{(m)} & \colhead{(deg)}    & \colhead{$V_c$} & \colhead{$\sigma V_c$} \\ }
\startdata
16141 & 54718.438 & 285.47 & 237.1 & 0.915 & 0.134 \\
      & 54718.445 & 281.55 & 238.2 & 0.944 & 0.123 \\
      & 54718.450 & 278.72 & 239.0 & 0.845 & 0.113 \\
      & 54718.455 & 275.69 & 240.0 & 0.961 & 0.100 \\
      & 54718.461 & 272.09 & 241.2 & 0.815 & 0.104 \\
      & 54718.467 & 269.11 & 242.2 & 0.834 & 0.115 \\
      & 54718.472 & 266.31 & 243.3 & 0.900 & 0.119 \\
      & 54718.477 & 263.67 & 244.4 & 0.972 & 0.143 \\
      & 54718.482 & 260.84 & 245.6 & 0.845 & 0.146 \\
      & 54718.487 & 258.06 & 246.8 & 0.977 & 0.138 \\

17092 & 54720.344 & 285.50 & 221.5 & 0.904 & 0.107 \\
      & 54720.354 & 290.19 & 223.3 & 0.841 & 0.093 \\
      & 54720.364 & 295.06 & 225.3 & 0.811 & 0.081 \\
      & 54720.371 & 297.63 & 226.5 & 0.753 & 0.079 \\
      & 54720.380 & 301.41 & 228.4 & 0.846 & 0.075 \\

45410 & 54720.481 & 258.11 & 212.9 & 0.696 & 0.078 \\
      & 54720.490 & 263.24 & 215.0 & 0.651 & 0.053 \\
      & 54720.496 & 266.69 & 216.4 & 0.587 & 0.073 \\
      & 54720.502 & 269.68 & 217.7 & 0.665 & 0.106 \\
      & 54720.509 & 272.90 & 219.2 & 0.716 & 0.097 \\

154345 & 54719.168 & 328.79 & 90.5 & 0.885 & 0.094 \\
       & 54719.179 & 328.73 & 93.3 & 0.843 & 0.109 \\
       & 54719.185 & 328.66 & 94.7 & 0.811 & 0.089 \\
       & 54719.192 & 328.57 & 96.2 & 0.803 & 0.096 \\
       & 54719.198 & 328.45 & 97.6 & 0.847 & 0.096 \\
       & 54719.204 & 328.29 & 99.2 & 0.903 & 0.095 \\
       & 54719.213 & 328.00 & 101.4 & 0.817 & 0.122 \\

185269 & 54665.204 & 321.00 & 228.6 & 0.860 & 0.146 \\
       & 54665.216 & 323.97 & 230.0 & 0.946 & 0.129 \\
       & 54665.226 & 326.17 & 231.3 & 0.757 & 0.148 \\
       & 54665.236 & 327.81 & 232.6 & 0.926 & 0.110 \\
       & 54665.245 & 328.96 & 233.9 & 0.928 & 0.178 \\
       & 54665.404 & 323.06 & 266.1 & 0.771 & 0.064 \\
       & 54665.410 & 322.92 & 267.7 & 0.741 & 0.050 \\
       & 54665.417 & 322.85 & 269.2 & 0.816 & 0.048 \\
       & 54665.423 & 322.85 & 90.8 & 0.921 & 0.057 \\
       & 54665.430 & 322.93 & 92.4 & 0.877 & 0.075 \\
       & 54665.438 & 323.11 & 94.3 & 0.912 & 0.084 \\
       & 54665.445 & 323.35 & 96.0 & 0.910 & 0.091 \\
       & 54665.452 & 323.68 & 97.7 & 0.855 & 0.080 \\
       & 54665.459 & 324.06 & 99.4 & 0.927 & 0.083 \\
       & 54665.466 & 324.52 & 101.1 & 0.841 & 0.129 \\
       & 54667.381 & 323.73 & 262.0 & 1.004 & 0.096 \\
       & 54667.387 & 323.44 & 263.5 & 0.830 & 0.103 \\
       & 54667.393 & 323.21 & 264.9 & 0.892 & 0.096 \\
       & 54667.400 & 323.02 & 266.5 & 1.014 & 0.085 \\
       & 54667.406 & 322.90 & 267.9 & 0.899 & 0.113 \\

188310 & 54717.211 & 293.54 & 249.5 & 0.103 & 0.014 \\
       & 54717.223 & 289.87 & 252.2 & 0.106 & 0.017 \\
       & 54717.229 & 288.10 & 253.7 & 0.107 & 0.012 \\
       & 54717.236 & 286.11 & 255.5 & 0.106 & 0.014 \\
       & 54717.242 & 284.72 & 257.0 & 0.094 & 0.015 \\
       & 54717.248 & 283.37 & 258.5 & 0.110 & 0.019 \\
       & 54717.253 & 282.29 & 260.0 & 0.111 & 0.018 \\
       & 54717.259 & 281.25 & 261.6 & 0.127 & 0.018 \\

199665 & 54717.336 & 285.96 & 90.6 & 0.614 & 0.064 \\
       & 54717.341 & 286.09 & 92.1 & 0.567 & 0.062 \\
       & 54717.347 & 286.36 & 93.6 & 0.562 & 0.077 \\
       & 54717.352 & 286.78 & 95.0 & 0.574 & 0.053 \\
       & 54717.358 & 287.37 & 96.6 & 0.566 & 0.060 \\
       & 54717.364 & 288.15 & 98.2 & 0.512 & 0.055 \\
       & 54717.370 & 289.11 & 99.1 & 0.479 & 0.069 \\
       & 54717.377 & 290.31 & 101.5 & 0.482 & 0.049 \\
       & 54717.383 & 291.58 & 103.1 & 0.414 & 0.035 \\
       & 54717.390 & 293.30 & 104.9 & 0.500 & 0.065 \\

210702 & 54717.426 & 302.96 & 100.6 & 0.635 & 0.076 \\
       & 54717.436 & 304.66 & 103.1 & 0.652 & 0.072 \\
       & 54717.442 & 305.68 & 104.5 & 0.591 & 0.085 \\
       & 54717.448 & 306.87 & 105.9 & 0.640 & 0.091 \\

217107 & 54717.283 & 292.41 & 236.3 & 0.771 & 0.096 \\
       & 54717.289 & 289.09 & 237.2 & 0.793 & 0.127 \\
       & 54717.296 & 285.35 & 238.3 & 0.757 & 0.095 \\
       & 54717.303 & 281.40 & 239.5 & 0.799 & 0.118 \\
       & 54717.309 & 278.11 & 240.6 & 0.776 & 0.114 \\

221345 & 54720.234 & 313.74 & 229.1 & 0.278 & 0.031 \\
       & 54720.239 & 315.41 & 229.9 & 0.253 & 0.034 \\
       & 54720.245 & 317.13 & 230.8 & 0.266 & 0.028 \\
       & 54720.250 & 318.64 & 231.7 & 0.232 & 0.024 \\
       & 54720.256 & 320.12 & 232.7 & 0.251 & 0.028 \\

222404 & 54664.457 & 253.07 & 230.4 & 0.105 & 0.011 \\
       & 54664.466 & 254.63 & 233.0 & 0.099 & 0.011 \\
       & 54664.475 & 256.07 & 235.6 & 0.091 & 0.010 \\
       & 54720.278 & 247.87 & 222.5 & 0.104 & 0.012 \\
       & 54720.285 & 249.26 & 224.5 & 0.093 & 0.010 \\	
       & 54720.295 & 251.32 & 227.6 & 0.093 & 0.008 \\
       & 54720.301 & 252.45 & 229.3 & 0.086 & 0.008 \\
       & 54720.307 & 253.58 & 231.2 & 0.092 & 0.009 \\
       & 54720.313 & 254.70 & 233.2 & 0.091 & 0.008 \\
       & 54720.320 & 255.83 & 235.2 & 0.087 & 0.009 \\
\enddata
\tablecomments{The projected baseline position angle ($\Theta$) is calculated to be east of north.}
\end{deluxetable}

\clearpage

%%%%%%%%%%%%%%%%%%%%%%%%%%%%% Diameter Measurements %%%%%%%%%%%%%%%%%%%%%%%%%%%%%%%%%%%%%%%%

\begin{deluxetable}{cccccccccc}
%\rotate
\tablewidth{1.0\textwidth}
\tabletypesize{\scriptsize}
\tablecaption{Exoplanet Host Star Angular Diameters and Radii.\label{calculating_diam}}

\tablehead{ \colhead{ } & \colhead{Spectral} & \colhead{ } & \colhead{$\pi$} & \colhead{$\theta_{\rm SED}$} & \colhead{$\theta_{\rm UD}$} & \colhead{$\theta_{\rm LD}$} & \colhead{$\sigma_{\rm LD}$} & \colhead{$R_{\rm L}$} &\colhead{$\sigma_{\rm R}$} \\
\colhead{HD} & \colhead{Type} & \colhead{$\mu_\lambda$} &\colhead{(mas)} & \colhead{(mas)} & \colhead{(mas)} & \colhead{(mas)} & \colhead{($\%$)} & \colhead{($R_\odot$)} &\colhead{($\%$)} \\ }
\startdata
16141 & G5 IV & 0.27 & 25.67$\pm$0.66 & 0.381$\pm$0.012$^\dagger$ & 0.480$\pm$0.048 & 0.490$\pm$0.049 & 10 & 2.05$\pm$0.21 & 10 \\
17092 & K0 III & 0.33 & (183$\pm$18 pc)$^\star$ & 0.531$\pm$0.029$^\dagger$ & 0.586$\pm$0.039 & 0.601$\pm$0.041 & 7 & 11.8$\pm$1.4 & 12 \\
45410 & K0 III-IV & 0.31 & 17.92$\pm$0.47 & 0.867$\pm$0.066 & 0.946$\pm$0.034 & 0.970$\pm$0.035 & 4 & 5.82$\pm$0.26 & 4 \\
154345 & G8 V & 0.28 & 53.80$\pm$0.32 & 0.452$\pm$0.008$^\dagger$ & 0.490$\pm$0.026 & 0.502$\pm$0.026 & 5 & 1.00$\pm$0.05 & 5 \\
185269 & G0 IV & 0.25 & 19.89$\pm$0.56 & 0.359$\pm$0.012$^\dagger$ & 0.471$\pm$0.032 & 0.480$\pm$0.033 & 7 & 2.59$\pm$0.19 & 7 \\
188310 & G9 III & 0.32 & 17.77$\pm$0.29 & 1.712$\pm$0.053 & 1.671$\pm$0.008 & 1.726$\pm$0.008 & 0.4 & 10.45$\pm$0.18 & 2 \\
199665 & G6 III & 0.31 & 13.28$\pm$0.31 & 0.985$\pm$0.028 & 1.083$\pm$0.027 & 1.111$\pm$0.028 & 3 & 9.00$\pm$0.31 & 3 \\
210702 & K1 III & 0.31 & 18.20$\pm$0.39 & 0.879$\pm$0.049$^\dagger$ & 0.854$\pm$0.017 & 0.875$\pm$0.018 & 2 & 5.17$\pm$0.15 & 3 \\
217107 & G8 IV & 0.28 & 50.36$\pm$0.38 & 0.534$\pm$0.016$^\dagger$ & 0.688$\pm$0.013 & 0.704$\pm$0.013 & 2 & 1.50$\pm$0.03 & 2 \\
221345 & G8 III & 0.32 & 12.63$\pm$0.27 & 1.380$\pm$0.164 & 1.297$\pm$0.008 & 1.336$\pm$0.009 & 1 & 11.38$\pm$0.26 & 2 \\
222404 & K1 IV & 0.32 & 70.91$\pm$0.40 & 3.130$\pm$0.211 & 3.331$\pm$0.022 & 3.302$\pm$0.029 & 1 & 5.01$\pm$0.05 & 1 \\
\enddata
\tablecomments{$^\star$HD~17092 had no parallax measurements available so we used the distance estimate from \citet{2008AstL...34..785G}. \\
All spectral classes are from the \emph{SIMBAD Astronomical Database}; $\mu_\lambda$ values are from \citet{1995A&AS..114..247C}; $\pi$ values are from \citet{2007hnrr.book.....V}. \\
$^\dagger$$\theta_{\rm SED}$ from \citet{2009ApJ...694.1085V}; otherwise SEDs were completed using photometry from the following sources: HD 45410: $UBV$ from \citet{1966CoLPL...4...99J}, $RI$ from \citet{2003AJ....125..984M}; HD 188310: $UBVRI$ from \citet{1978A&AS...34..477M}; HD 199665: $BV$ from \citet{1997ESASP1200.....P}, $RI$ from \citet{2003AJ....125..984M}; HD 221345: $UBV$ from \citet{1966CoLPL...4...99J}, $RI$ from \citet{2003AJ....125..984M}; and HD 222404: $UBVRI$ from \citet{1978A&AS...34..477M}. All $JHK$ values from \citet{2003tmc..book.....C}.}
\end{deluxetable}

\clearpage

%%%%%%%%%%%%%%%%%%%%%%%%%%%%% All Calibrated Visibilities %%%%%%%%%%%%%%%%%%%%%%%%%%%%%%%%%%%

\begin{deluxetable}{llll}
\tablewidth{0pc}
\tablecaption{Binary and Variable Stars in the Sample.\label{binvar}}

\tablehead{
 \colhead{Target} & \colhead{ }  & \colhead{ }   & \colhead{ }  \\
 \colhead{HD}   & \colhead{Type} & \colhead{Reference} & \colhead{Notes} \\ }
\startdata
16141 & binary & Mugrauer et al. (2005) & $\rho$ = 6 arcsec; outside Array's FOV$^\dagger$ \\
45410 & binary & \citet{2001AJ....122.3466M} & $\rho$ = 190 arcsec; outside Array's FOV$^\dagger$ \\
154345 & variable & \citet{2009yCat....102025S} & no variability period or type listed \\
185269 & binary & \citet{1989ApJS...69..141S} & listed as binary but no orbital info given; \\
 & & & no other indication in literature of binarity \\
188310 & binary & \citet{2001AJ....122.3466M} & $\rho$ = 0.1 arcsec, $\Delta$m$_V$ =4.7; outside range of Array \\
199665 & binary & \citet{2001AJ....122.3466M} & $\rho$ = 130 - 200 arcsec; outside Array's FOV$^\dagger$  \\
217107 & binary & \citet{2001AJ....122.3466M} & $\rho$ = 0.3 - 0.5 arcsec; outside Array's FOV$^\dagger$ \\
221345 & variable & \citet{1982bscf.book.....H} & no variability detected in \citet{1993PASP..105.1422P} \\
222404 & binary & \citet{2007ApJ...654.1095T} & $\rho$ = 325 mas, $\Delta$m$_K$ = 6.4; outside range of Array \\
\enddata
\tablecomments{$\rho$ = binary separation, $\Delta$m = magnitude difference \\
$^\dagger$The field of view depends largely on the baseline used in the observations, so while some of the secondary companions would affect the data on shorter baselines, they will not be visible in the measurements on the baseline used here.}
\end{deluxetable}

\clearpage

%%%%%%%%%%%%%%%%%%%%%%%%%%%%% Effective Temperatures %%%%%%%%%%%%%%%%%%%%%%%%%%%%%%%%%%%%%%%%

\begin{deluxetable}{ccrclccr}
%\rotate
\tablewidth{0pc}
\tabletypesize{\scriptsize}
\tablecaption{Stellar Effective Temperatures and Luminosities.\label{temps}}

\tablehead{ \colhead{Star} & \colhead{$A_{\rm V}$} & \colhead{  } & \colhead{$F_{\rm BOL}$}            & \colhead{Calculated}        & \colhead{$\sigma_{T \rm eff}$} & \colhead{Range of $T_{\rm eff}$ from} & \colhead{log($L$)}  \\
           \colhead{HD}    & \colhead{(mag)}     &  \colhead{BC} & \colhead{(10$^{-8}$ erg s$^{-1}$ cm$^{-2}$)} & \colhead{$T_{\rm eff}$ (K)} & \colhead{($\%$)}           & \colhead{other sources (K)}  & \colhead{($L_{\odot}$)} \\ }
\startdata
16141 & 0.00$^{\rm a}$ & 0.06$\; \pm \;$0.04 & 4.9$\; \pm \;$0.2 & 4982$\; \pm \;$254 & 5 & 4900-5888 & 2.3$\; \pm \;$0.1  \\
17092 & 0.80$^{\rm a}$ & 0.50$\; \pm \;$0.05 & 6.2$\; \pm \;$0.4 & 4765$\; \pm \;$182 & 4 & 4750 & 65.0$\; \pm \;$3.1 \\
45410 & 0.03$^{\rm b}$ & 0.29$\; \pm \;$0.03 & 15.2$\; \pm \;$0.5 & 4689$\; \pm \;$92 & 2 & 4750-4898 & 14.8$\; \pm \;$0.4 \\
154345 & 0.20$^{\rm a}$ & 0.40$\; \pm \;$0.04 & 8.6$\; \pm \;$0.4 & 5664$\; \pm \;$158 & 3 & 5436-5570 & 0.9$\; \pm \;$0.0 \\
185269 & 0.13$^{\rm a}$ & 0.01$\; \pm \;$0.03 & 6.0$\; \pm \;$0.2 & 5283$\; \pm \;$186 & 4 & 5850-6166 & 4.7$\; \pm \;$0.1 \\
188310 & 0.10$^{\rm b}$ & 0.35$\; \pm \;$0.02 & 50.2$\; \pm \;$1.0 & 4742$\; \pm \;$26 & 1 & 4635-4786 & 49.7$\; \pm \;$0.9 \\
199665 & 0.00$^{\rm b}$ & 0.28$\; \pm \;$0.04 & 26.8$\; \pm \;$1.1 & 5054$\; \pm \;$81 & 2  & 4750-5012 & 47.6$\; \pm \;$1.8 \\
210702 & 0.10$^{\rm a}$ & 0.32$\; \pm \;$0.03 & 14.2$\; \pm \;$0.4 & 4859$\; \pm \;$62 & 1 & 4600-4898 & 13.4$\; \pm \;$0.4 \\
217107 & 0.10$^{\rm a}$ & 0.09$\; \pm \;$0.03 & 9.5$\; \pm \;$0.3 & 4895$\; \pm \;$57 & 1 & 4900-5704 & 1.2$\; \pm \;$0.0 \\
221345 & 0.13$^{\rm b}$ & 0.36$\; \pm \;$0.03 & 32.3$\; \pm \;$1.0 & 4826$\; \pm \;$40 & 1 & 4582-4900 & 63.3$\; \pm \;$1.8 \\
222404 & 0.01$^{\rm b}$ & 0.36$\; \pm \;$0.00 & 184.0$\; \pm \;$0.5 & 4744$\; \pm \;$21 & 0.4 & 4566-4916 & 11.4$\; \pm \;$0.0 \\
\enddata
\tablecomments{$^{\rm a}$\citet{2009ApJ...694.1085V}; $^{\rm b}$\citet{2005A&A...430..165F}. \\ 
All BC values from \citet{1999A&A...352..555A} except for HD 17092 and HD 154345, which are from \citet{2000asqu.book.....C} with an assigned error of 10$\%$. \\
The range of $T_{\rm eff}$ values are from the \emph{VizieR database of astronomical catalogs} \citep{2000A&AS..143...23O}.}
\end{deluxetable}

\clearpage

%%%%%%%%%%%%%%%%%%%%%%%%%%%%% Model Outputs %%%%%%%%%%%%%%%%%%%%%%%%%%%%%%%%%%%%%%%%

\begin{deluxetable}{ccrrcc}
%\rotate
\tablewidth{0pc}
%\tabletypesize{\scriptsize}
\tablecaption{PARAM Model Results.\label{models}}

\tablehead{\colhead{Target} & \colhead{$V$} & \colhead{Average} & \colhead{$R_{\rm model}$} & \colhead{Mass}         & \colhead{Age}   \\
           \colhead{HD}   & \colhead{mag} & \colhead{[Fe/H]}  & \colhead{($R_{\odot}$)}   & \colhead{($M_{\odot}$)} & \colhead{(Gyr)} \\ }
\startdata
16141 & 6.83 & 0.11$\; \pm \;$0.07 & 2.3$\; \pm \;$0.1 & 1.1$\; \pm \;$0.0 & 7.2$\; \pm \;$1.1 \\
17092 & 7.82 & 0.00$\; \pm \;$0.05 & 7.8$\; \pm \;$0.4 & 1.5$\; \pm \;$0.2 & 2.6$\; \pm \;$0.9 \\
45410 & 5.87 & 0.17$\; \pm \;$0.05 & 6.1$\; \pm \;$0.3 & 1.3$\; \pm \;$0.1 & 4.0$\; \pm \;$1.3 \\
185269 & 6.70 & 0.11$\; \pm \;$0.05 & 2.6$\; \pm \;$0.1 & 1.4$\; \pm \;$0.0 & 3.4$\; \pm \;$0.2 \\
188310 & 4.70 & -0.27$\; \pm \;$0.10 & 10.0$\; \pm \;$0.4 & 1.0$\; \pm \;$0.2 & 7.1$\; \pm \;$3.6 \\
199665 & 5.48 & -0.10$\; \pm \;$0.12 & 8.0$\; \pm \;$0.3 &  2.0$\; \pm \;$0.1 & 1.1$\; \pm \;$0.1 \\
210702 & 5.95 & 0.00$\; \pm \;$0.05  & 5.2$\; \pm \;$0.2 & 1.4$\; \pm \;$0.1 & 3.5$\; \pm \;$1.1 \\
221345 & 5.22 & -0.32$\; \pm \;$0.05 & 10.3$\; \pm \;$0.3 & 1.1$\; \pm \;$0.2 & 4.5$\; \pm \;$1.9 \\
222404 & 3.21 & 0.08$\; \pm \;$0.11 & 5.0$\; \pm \;$0.2 & 1.2$\; \pm \;$0.1 & 5.4$\; \pm \;$2.1 \\
\enddata
\tablecomments{$V$ magnitudes are from Mermilliod (1991) except for HD~17092, which is from \citet{2006PASP..118.1666D}; Average [Fe/H] are from the literature; $R_{\rm model}$, Mass, and Age are model outputs.}
\end{deluxetable}

\clearpage

\clearpage

%%%%%%%%%%%%%%%%%%%%%%%%%%%%%%%%%%%%%% LD Plots %%%%%%%%%%%%%%%%%%%%%%%%%%%%%%%%%%%%%%%%%%%

\clearpage

\clearpage

%%%%%%%%%%%%%%%%%%%%%%%%%%%%%%%%%%%%% SED vs LD %%%%%%%%%%%%%%%%%%%%%%%%%%%%%%%%%%%%%%%%%%%

\clearpage

%%%%%%%%%%%%%%%%%%%%%%%%%%%%%%%%%%%%% Radii Compare %%%%%%%%%%%%%%%%%%%%%%%%%%%%%%%%%%%%%%%%%%%

%%%%%%%%%%%%%%%%%%%%%%%%%%%%%%%%%%%%% Radii Compare %%%%%%%%%%%%%%%%%%%%%%%%%%%%%%%%%%%%%%%%%%%

\end{document}